\documentclass[aoas,MSNbibl,nameyear,seceqn,dvips]{arximspdf}
\usepackage{graphicx}
\doi{10.1214/13-AOAS644} 
\volume{7}
\issue{3}
\pubyear{2013}
\firstpage{1709}
\lastpage{1732}



\begin{document}
\begin{frontmatter}

\title{Maximum likelihood estimation of mark--recapture--recovery models
in the presence~of~continuous covariates}
\runtitle{ML estimation for MRR data with continuous covariates}

\begin{aug}
\author[A]{\fnms{Roland} \snm{Langrock}\corref{}\thanksref{t1}\ead[label=e1]{roland@mcs.st-and.ac.uk}}
\and
\author[A]{\fnms{Ruth} \snm{King}\ead[label=e2]{ruth@mcs.st-and.ac.uk}}
\runauthor{R. Langrock and R. King}
\affiliation{University of St Andrews}
\address[A]{Centre for Research into Ecological\\
\quad and Environmental Modelling\\
School of Mathematics and Statistics \\
The Observatory \\
Buchanan Gardens \\
University of St Andrews \\
St Andrews \\
KY16 9LZ \\
United Kingdom \\
\printead{e1}\\
\hphantom{E-mail: }\printead*{e2}} 
\end{aug}

\thankstext{t1}{Supported by the Engineering and Physical Sciences Research
Council.}

\received{\smonth{11} \syear{2012}}
\revised{\smonth{1} \syear{2013}}

%
\begin{abstract}
We consider mark--recapture--recovery (MRR) data of
animals where the model parameters are a function of individual
time-varying continuous covariates. For such covariates, the covariate
value is unobserved if the corresponding individual is unobserved, in
which case the survival probability \mbox{cannot} be evaluated. For
continuous-valued covariates, the corresponding likelihood can only be
expressed in the form of an integral that is analytically intractable
and, to date, no maximum likelihood approach that uses all the
information in the data has been developed. Assuming a first-order
Markov process for the covariate values, we accomplish this task by
formulating the MRR setting in a state-space framework and considering
an approximate likelihood approach which essentially discretizes the
range of covariate values, reducing the integral to a summation. The
likelihood can then be efficiently calculated and maximized using
standard techniques for hidden Markov models. We initially assess the
approach using simulated data before applying to real data relating to
Soay sheep, specifying the survival probability as a function of body
mass. Models that have previously been suggested for the corresponding
covariate process are typically of the form of diffusive random walks.
We consider an alternative nondiffusive AR(1)-type model which appears
to provide a significantly better fit to the Soay sheep data. 
\end{abstract}

%
\begin{keyword}
\kwd{Arnason--Schwarz model}
\kwd{hidden Markov model}
\kwd{Markov chain}
\kwd{missing values}
\kwd{Soay sheep}
\kwd{state-space model}
\end{keyword}

\end{frontmatter}

\section{Introduction}
\label{intro}

Mark--recapture--recovery (MRR) data are commonly collected on animal
populations in order to gain some understanding of the underlying
system. Data are collected by repeated surveyings of the population
under study. In the initial survey all individuals that are observed
are uniquely identified (via natural features or by applying some form
of mark, such as a ring or tag) and released back into the population.
At each subsequent survey all individuals observed are recorded, and
those that have not previously been observed are again uniquely
identified, before all are released back into the population. We assume
that individuals can be observed alive or recovered dead in each
survey. The resulting MRR data can be summarised as the observed
encounter histories for each individual observed within the population,
detailing for each survey event whether an individual was observed
alive or recovered dead. Conditioning on the initial capture time of
each individual leads to Cormack--Jolly--Seber-type models [see
\citet{schs00} for a review of these models]. The original
Cormack--Jolly--Seber model considered only live captures (i.e.,
mark--recapture data) and was extended to additional recoveries by
\citet{bar97}. The corresponding MRR likelihood function of these data
can be written as a function of survival, recapture and recovery
probabilities.

Recent research has focussed on linking environmental and individual
covariates to demographic parameters, most notably the survival
probabilities, in order to explain temporal and individual variability
[\citet
{brocm00,catmcfa00,cou01,pol02,kinb03,kinbmc06,gimcbjm06,catmt08,schb11},
to name but a few]. We consider individual time-varying continuous
covariates. These have traditionally been difficult to deal with due to
the missing covariate values (if an individual is unobserved, the
corresponding covariate value is also unknown). 
One of the initial approaches to dealing with such covariates was to
(coarsely) discretize the covariate space, essentially defining
discrete covariate ``states.'' \citet{nicsph92} considered data relating
to meadow voles (\emph{Microtus pennsylvanicus}) and categorised weight
into four different categories. Such a discretization reduces the model
to the Arnason--Schwarz model [\citet{brownie93,schwarz93}]. Transition
probabilities between the covariate states are estimated within the
optimisation of the likelihood (possibly with additional restrictions
on the state transitions). With the coarse discretization arbitrarily
defined, this approach leads to a (potentially significant) loss of
information. \citet{catmt08} have proposed a conditional likelihood
approach (often referred to as the ``trinomial approach''). By
conditioning on only the observed covariate values, this approach
results in a simple, closed-form likelihood expression. However, this
involves discarding a proportion of the available data, leading to a
decreased precision of the parameter estimates. In addition, Bayesian
approaches have been proposed [\citet{bons06,kinbc08}] and the
corresponding model fitted using a data augmentation approach [\citet
{tanw87}]. 
Within the Bayesian approach priors need to specified on the model
parameters (and possibly models in the presence of model uncertainty).
In addition, model selection is generally more difficult due to
computational complexity, and posterior model probabilities can be
sensitive to the prior distributions specified on the parameters. See
\citet{bonmk10} for further discussion and a comparison of the Bayesian
and trinomial approaches, and \citet{catmt08} for an overview of the
existing approaches.

For the considered type of MRR data, \citet{bonmk10} state that ``except
when few values are missing, the large number of integrals [\ldots] will
make it impossible to perform maximum likelihood estimation'' (page
1258). We claim that this statement is not true and present a novel
approach based on a hidden Markov-type formulation of the MRR setting.
This formulation leads to a likelihood that is easy to compute and to
maximize numerically. The underlying idea is to finely discretize the
space of possible covariate values, which corresponds to a numerical
integration of the likelihood function. The numerical integration
enables us to augment the resulting discrete space of covariate values
with the state space of the survival process, leading to a single,
partially hidden Markov process for each observed encounter history.
This approach essentially extends the previous coarse discretization
approach of \citet{nicsph92} by considering a very fine discretization
of the covariate space, coupled with specifying structured transition
probability matrices defined using a covariate process model. The
corresponding likelihood can be written in a closed and efficient
matrix product form that is characteristic of hidden Markov models
(HMMs) [\citet{zucm09}]. 
Notably, model selection can be carried out using standard model
selection techniques.

We apply the method to data relating to Soay sheep (\textit{Ovis aries}).
The Soay sheep on the uninhabitated island of Hirta in the St Kilda
archipelago, Scotland, are a well-studied biological system [\citet
{clup04}]. Intensive annual surveys involve physical recaptures of
individuals, tagging of lambs, visual resightings and searches for dead
carcasses. A range of individual covariate data are recorded for each
sheep. We focus on the body mass recorded, collected (when possible)
when an individual is physically recaptured. Males and females have
different life strategies, and we consider data relating to only
females, tagged as lambs between 1985--2008 and recaptured/recovered
annually from 1986--2009. We investigate the effect of body mass on
survival and consider a variety of models for the change of body mass
over time. The latter aspect is usually not the primary focus of MRR
studies, although it is clearly of biological interest. In particular,
we demonstrate that the (diffusive) models that have previously been
considered for the change of body mass over time are outperformed by
alternative (nondiffusive) AR(1)-type processes.

The manuscript is structured as follows. Section~\ref{sect2} introduces
the HMM-type estimation method for the specific MRR setting under
consideration. An extensive simulation study investigating the
performance of the proposed method, including a comparison to the
trinomial approach, is given in Section~\ref{simul}. In Section~\ref{appl} we analyse MRR data collected on Soay sheep, where the
time-varying covariate of interest corresponds to body mass. We
conclude with a discussion in Section~\ref{discuss}.

\section{Hidden Markov-type formulation of the MRR setting}\label{sect2}

We initially develop the form of the (partially) hidden Markov model
for standard MRR data (i.e., without any covariate information, in
Section~\ref{noco}), before extending to include individual
time-varying continuous covariate information (in Section~\ref{onco}).

\subsection{Formulation in absence of covariate information}\label{noco}

\subsubsection{General model formulation and notation}

MRR data are typically most easily expressed in the form of the capture
history of each individual animal observed within the study. We
initially consider the probability of an encounter history for a given
individual. Suppose that there are $T$ capture occasions within the
study. The capture history for the individual is denoted by
$(x_1,\ldots,x_T)$, such that
\[
x_{t} = \cases{1, &\quad if the individual is observed at time $t$;
\cr
2,
&\quad if the individual is recovered dead in the interval $(t-1, t]$;
\cr
0, &\quad
otherwise.}
\]
Following the initial capture of the individual, the encounter history
can be regarded as the combination of two distinct processes: an
underlying survival process and an observation process, conditional on
the survival state of an individual. Thus, MRR data can be modelled via
a (discrete) state-space model (i.e., HMM), separating the underlying
state process (i.e., survival process) from the observation process
(i.e., recapture/recovery processes). For further discussion we refer
the reader to \citet{gimrcddvvp07}, \citet{schb08}, \citet{royle08}, \citet
{kinmgb09} and \citet{kin12}. Let $g$ denote the occasion on which the
individual is initially observed and marked.
We define the survival process, $(s_g,\ldots,s_T)$, such that
\[
s_{t} = \cases{1, &\quad if the individual is alive at time $t$;
\cr
2, &\quad
if the individual is dead at time $t$, but was alive at time $t-1$;
\cr
3, &\quad if
the individual is dead at time $t$, and was dead at time $t-1$.}
\]
Note that here we explicitly distinguish between ``recently dead''
individuals ($s_{t} = 2$) and ``long dead'' individuals ($s_{t}=3$), and
assume that only recently dead individuals can be recovered dead at a
given capture event. This is a standard assumption within MRR models,
due to the decay of marks for identifying individuals once they have
died [although see, e.g., \citet{CFMN01}, where this assumption is not valid].

The likelihood of the observed capture histories is a function of
survival, recapture and recovery probabilities. In particular, we set
\begin{eqnarray*}
\phi_{t} &=& P( s_{t+1}=1 \vert s_{t}=1) \qquad\mbox{(\textit{survival probability})},
\\
p_{t} &=& P( x_{t}= 1 \vert s_{t}=1) \qquad\mbox{(\textit{capture probability})},
\\
\lambda_{t} &=& P( x_{t}= 2 \vert s_{t}=2)\qquad
\mbox{(\textit{recovery probability})}.
\end{eqnarray*}
We note that the survival process is only partially observed (i.e., it
is partially hidden). For a capture history that includes a dead
recovery, the corresponding survival process is completely known
following initial capture (i.e., if $x_{\tau} = 2$, then $s_{t} = 1$
for $t=g,\ldots,\tau-1$, $s_{t}=2$ for $t=\tau$ and $s_{t} = 3$ for
$t=\tau+1,\ldots,T$). Similarly, if an individual is observed at the
final capture event, then the associated survival process following
initial capture is also fully known (i.e., if $x_{T} = 1$, then $s_{t}
= 1$ for $t=g,\ldots,T$). However, for all other histories the survival
process following the final capture of the individual is unknown. For
notational convenience, we let $\mathcal{S} = \{ t\geq g\dvtx  s_{t}$
is known$\}$ denote the set of all occasions at which the survival
state of the individual is known, and $\mathcal{S}^c$ the corresponding
complement, that is, the set of occasions at which the survival state
is unknown, following initial capture.

\subsubsection{The likelihood}\label{seclic1}

Conditional on the initial capture, the likelihood for a single capture
history can be written in the form
%
\begin{equation}
\label{lik1} \mathcal{L} = \sum_{\tau\in\mathcal{S}^c} \sum
_{s_{\tau} \in\{
1,2,3\}} \prod_{t=g+1}^T
f(s_{t}\vert s_{t-1}) f(x_{t}\vert
s_{t}),
\end{equation}
taking into account all possible survival histories for the animal,
given its observed capture history. For notational simplicity, we use
$f$ as a general symbol for a probability mass function or a density
function, possibly conditional, throughout the manuscript. For example, here
%
\[
f(s_t | s_{t-1}) = \cases{ \phi_{t-1}, &\quad
$s_t = 1; s_{t-1} = 1$;
\cr
1 - \phi_{t-1}, &\quad
$s_t = 2; s_{t-1} = 1$;
\cr
1, &\quad $s_t = 3;
s_{t-1} \in\{ 2, 3\}$;
\cr
0, &\quad otherwise,}
\]
and
\[
f(x_t | s_t) = \cases{p_t, &\quad
$s_t = 1; x_t = 1$;
\cr
1 - p_t, &\quad
$s_t = 1; x_t = 0$;
\cr
\lambda_t, &\quad
$s_t = 2; x_t = 2$;
\cr
1-\lambda_t, &\quad
$s_t = 2; x_t = 0$;
\cr
1, &\quad $s_t = 3;
x_t = 0$;
\cr
0, &\quad otherwise.}
\]
Expression (\ref{lik1}) represents an inefficient way of computing the
likelihood, since some impossible state sequences are taken into
account (such as, e.g., $\ldots,1,2,1,1,\ldots$) that have a zero
contribution to the likelihood. Clearly, only possible state sequences
need to be evaluated, but we retain the full summation for notational
simplicity.

An alternative expression for the likelihood is available using matrix
products. In particular, at time $t$, we define the transition
probability matrix associated with the transitions between different
survival states by $\bolds{\Gamma}_{t}$, such that
\[
\bolds{\Gamma}_{t} = \pmatrix{ \phi_{t} & 1-
\phi_{t} & 0
\cr
0 & 0 & 1
\cr
0 & 0 & 1}.
\]
Furthermore, let $\mathbf{Q}(x_{t})$ denote the diagonal matrix giving
the state-dependent probabilities of observations at time $t$ on the diagonal:
\[
\mathbf{Q}(x_{t}) = \cases{\operatorname{diag}(1-p_{t},1-
\lambda_{t},1), &\quad if $x_{t} = 0$;
\cr
\operatorname{diag}(p_{t},0,0),
&\quad if $x_{t} = 1$;
\cr
\operatorname{diag}(0,\lambda_{t},0), &\quad if
$x_{t} = 2$,}
\]
where diag$(\ldots)$ denotes the diagonal matrix with given diagonal
elements. The likelihood (\ref{lik1}) can then be written in the HMM form
%
\begin{eqnarray}
\label{lik2}
\mathcal{L} & = & \bolds{\delta} \Biggl( \prod
_{t=g+1}^{T} \bolds{\Gamma}_{t-1}
\mathbf{Q}(x_{t}) \Biggr) \mathbf {1}_3
\nonumber\\[-8pt]\\[-8pt]
& = & \bolds{\delta} \bolds{\Gamma}_{g}
\mathbf{Q}(x_{g+1}) \bolds{\Gamma}_{g+1}
\mathbf{Q}(x_{g+2}) \cdot\cdots\cdot \bolds{\Gamma}_{T-1}
\mathbf{Q}(x_{T})\mathbf{1}_3,\nonumber
\end{eqnarray}
where $\mathbf{1}_3$ denotes a column vector of length 3 with each
element equal to 1, and $\bolds{\delta}=(1,0,0)$ is the row vector
giving the conditional probabilities of occupying the different
survival states at the initial capture occasion, given that the
individual was captured. The likelihood (\ref{lik2}) is that of a
\textit{partially} hidden Markov model, and one effectively sums only over the
unknown states, rather than over all possible state sequences. We
further note that in general for MRR data, the likelihood can be
calculated more efficiently using sufficient statistics, but we
introduce this form of notation here for facilitating the extension to
time-varying individual covariates. In an MRR setting, the HMM-type
matrix product likelihood form has previously been given by \citet
{pra05}, who also discusses the general benefits of being able to apply
the powerful HMM machinery.

\subsection{Formulation in the presence of continuous-valued
covariates}\label{onco}

\subsubsection{General model formulation and notation}

We extend the HMM framework to allow for the inclusion of
individual-specific, continuous covariate\vadjust{\goodbreak} information that varies over
time. For example, this may correspond to the condition of the
individual (where proxies such as body mass or parasitic load may be
used). We consider a single time-varying continuous covariate, such
that the survival probabilities are a deterministic function of this covariate.
The extension of the method to multiple covariates is, in principle,
straightforward, although technically challenging and accompanied by
large scale increases in computational time (see Section~\ref{discuss}
for further discussion).

Notationally, for a given individual we let $y_{t}$ denote the value of
the covariate at time $t$, $t=g,\ldots,T$, and $\mathbf{y} = \{y_{t}\dvtx
t=g,\ldots,T\}$ the set of all covariate values. For all $t \ge\tau$
such that $x_{\tau} = 2$, the value of $y_{t}$ (i.e., the covariate
value following the observed death) is not defined. We note that
usually one observes $y_{t}$ when $x_{t} = 1$, but there may still be
cases where an individual is observed alive, but no covariate value is
recorded. This may occur, for example, due to a resighting rather than
a recapture of the individual, or time constraints making it infeasible
to obtain covariate values for all individuals observed. We let
$\mathcal{W}=\{t \geq g\dvtx  y_{t}$ is observed$\}$ denote the set of
times for which the covariate is observed. The corresponding observed
covariate values are denoted by $\mathbf{y}_{\mathcal{W}} = \{y_{t}\dvtx  t
\in\mathcal{W}\}$. Similarly, we let $\mathcal{W}^c$ denote the
complement, that is, the set of times for which the covariate is
unobserved, excluding times for which it is known the individual is not
in the study (i.e., before initial capture or when known to be dead),
so that $\mathcal{W}^c = \{t\geq g\dvtx  y_{t}$ is unobserved$\}
\setminus\{ t \geq g\dvtx  t \in\mathcal{S}, s_{t} = 2,3 \}$.
Finally, we let the set of missing covariate values be denoted by
$\mathbf{y}_{\mathcal{W}^c} = \{y_{t}\dvtx  t \in\mathcal{W}^c\}$.

We consider models in which the survival probability depends on the
covariate, and
assume that the probability of survival from occasion $t$ to $t+1$ is
determined by the value $y_{t}$. Typically a logistic regression of
survival probability on covariate value is considered, so that
%
\begin{equation}
\label{sumo} \operatorname{logit}(\phi_t) = \beta_0 +
\beta_1 y_{t};
\end{equation}
see, for example, \citet{norm79} and \citet{bonmk10}.

Following \citet{bonmk10}, we assume an underlying model for the change
in covariate values over time, specified by some first-order Markov
process, $f(y_{t}|y_{t-1})$, for $t=g+1,\ldots,T$. We set the function
value of $f(y_{t} | y_{t-1})$ to one for $s_{t}=2,3$ (i.e., when an
individual is dead). The covariate value may not be recorded at the
initial capture, in which case we also require an underlying
distribution on the initial covariate values, described by a
probability density function~$f_0$ (but see remarks at the end of
Section~\ref{liksec}). Typically a random walk-type model is assumed
for the underlying covariate model. For example, \citet{bons06} and \citet
{kinbc08} consider models along the lines of
%
\begin{equation}
\label{como} y_{t+1}|y_{t} \sim N\bigl(y_{t}+a_t,
\sigma^2\bigr)
\end{equation}
with $a_t$ varying over time, and extensions thereof to allow for
additional modelling complexities such as age-dependence. However,
fitting such models involves some complexities, due to the unobserved
covariate values, which need to be integrated out in order to
explicitly calculate the likelihood function of the data. We discuss
this in further detail next and propose a likelihood-based approach
that exploits the HMM machinery.

\subsubsection{The likelihood}\label{liksec}

With a first-order Markov process for the covariate values, the
likelihood of the capture history and observed covariate values of an
individual, conditional on the initial capture event, can be written in
the form
%
\begin{eqnarray}
\label{lik3} \mathcal{L} &=& \int\cdots \int\sum_{\tau\in\mathcal{S}^c}
\sum_{s_{\tau} \in\{ 1,2,3\}} f_0(y_g)
\nonumber\\[-8pt]\\[-8pt]
&&\hspace*{33pt}{} \times\prod_{t=g+1}^T
f(s_{t}\vert s_{t-1},y_{t-1}) f(x_{t}
\vert s_{t})f(y_{t}\vert y_{t-1}) \,d
\mathbf{y}_{\mathcal{W}^c}.\nonumber
\end{eqnarray}
In general, the necessary integration within this likelihood expression
is analytically intractable. In a Bayesian approach, the missing
covariate values are typically treated as auxiliary
variables that are essentially integrated out within the MCMC algorithm
[\citet{kinmgb09}]. However, model selection is usually complex in terms
of the estimation of the Bayes factors or posterior model probabilities
[although see \citet{kinbc08}, \citet{kinmgb09} and \citet{kibr02} with
regard to the use of the reversible jump (RJ)MCMC for covariate
selection and age/time dependence of the demographic parameters] and
the potential sensitivity of these on the prior specified on the model
parameters [see \citet{kinmgb09} for further discussion].

We adopt a classical maximum likelihood approach here, where we closely
approximate the multiple integral appearing in the likelihood using
numerical integration, essentially finely discretizing the space of
covariate values. This approach gives an approximation to the
likelihood which can be made arbitrarily accurate by increasing the
fineness of the discretization. In many MRR settings, the computational
effort required to obtain a very close approximation is very
reasonable, since one can evaluate the approximate likelihood using an
efficient HMM-type recursion (as shown below). The suggested strategy
for approximating the likelihood has previously been successfully
applied in finance in order to estimate stochastic volatility models
[see, e.g., \citet{frih98} and \citet{bar03}], but has a much wider scope
as pointed out by \citet{lan11}.

Mathematically, we define an ``essential range'' for the covariate values
and split this range into $m$ intervals of equal length, where $m$ is
some large number (e.g., $m=100$). Let the $j$th interval be denoted by
$B_j = [b_{j-1}, b_{j})$, $j=1,\ldots,m$. The essential range
corresponds to a lower and upper bound for the possible covariate
values, given by $b_0$ and $b_m$, respectively. We let $b_j^*$ denote a
representative point in~$B_j$. For large $m$ the choice of this point
only plays a very minor role, and throughout this manuscript we will
simply use the interval midpoint. The likelihood (\ref{lik3}) is then
approximated by
%
\begin{eqnarray}
\label{lik4} \mathcal{L}
&\approx& \sum_{\kappa\in\mathcal{W}^c} \sum
_{j_{\kappa
}=1}^m \sum_{\tau\in\mathcal{S}^c}
\sum_{s_{\tau} \in\{ 1,2,3\}} f_0(y_g)^{I_{\{g \in\mathcal{W}\}}}
\biggl(\int_{b_{j_g-1}}^{b_{j_g}} f_0(z)\,dz
\biggr)^{I_{\{g \in\mathcal{W}^c\}}}
\nonumber\\
&&{} \times\prod_{t=g+1}^T \bigl[
f(s_{t}\vert s_{t-1},y_{t-1})^{I_{\{ (t-1) \in\mathcal{W} \}}} f
\bigl(s_{t}\vert s_{t-1},b_{j_{t-1}}^*
\bigr)^{I_{\{ (t-1) \in\mathcal{W}^c \}}} f(x_{t} \vert s_{t})
\nonumber\\[-8pt]\\[-8pt]
&&\hspace*{39.6pt}{} \times f(y_{t} \vert y_{t-1})^{I_{\{(t-1) \in\mathcal{W},
t \in\mathcal{W} \}}} f
\bigl(y_{t} \vert b_{j_{t-1}}^*\bigr)^{I_{\{(t-1) \in
\mathcal{W}^c, t \in\mathcal{W} \}}}
\nonumber\\
&&\hspace*{39.6pt}{} \times f(y_{t} \in B_{j_t} \vert
y_{t-1})^{I_{\{(t-1) \in
\mathcal{W}, t \in\mathcal{W}^c \}}} f \bigl(y_{t} \in B_{j_t}
\vert b_{j_{t-1}}^*\bigr)^{I_{\{(t-1) \in\mathcal{W}^c, t \in\mathcal{W}^c \}}}
\bigr],\nonumber
\end{eqnarray}
where $I$ denotes the indicator function. In the last three lines in
(\ref{lik4}), the indicator function is used to distinguish between the
cases where the covariate value is known (so that the observed value
can be used) or unknown (so that the defined intervals and associated
representative values are used), at times $t-1$ and $t$. The final two
lines correspond to the likelihood contribution of the underlying model
for the covariate process and
%
\begin{equation}
\label{innerint} f (y_{t} \in B_{j} \vert z) = \int
_{b_{j-1}}^{b_{j}} f(y_t \vert z)
\,dy_t.
\end{equation}
Note this is essentially the same numerical integration strategy that
has previously been implemented by \citet{lan11} and \citet{lanmz12}; see
the latter reference for more details. The major difference to those
approaches is that here we allow for some covariate values to be
observed, and hence do not integrate over these observed covariate
values. Also, here there is the additional difficulty of a second level
of missing values, given by those $s_t$ with $t \in\mathcal{S}^c$,
which need to be summed over. Alternative numerical procedures for
evaluating the likelihood are discussed in Section~\ref{discuss}. In
cases where the integral appearing in (\ref{innerint}) cannot be solved
analytically, it can be approximated by $(b_j-b_{j-1}) f(b_j^* | z)$. %

The likelihood (\ref{lik4}) can be written in HMM-type matrix notation,
corresponding to an efficient recursive scheme for evaluating the
likelihood [see \citet{zuc08} for a more detailed description of the
recursion]. This makes maximum likelihood estimation feasible and has
the general benefit that the well-developed HMM machinery becomes
applicable. To do this, we essentially augment the ``alive'' survival
state by dividing it into $m$ distinct states, corresponding to ``alive
and with covariate value in $B_j$,'' $j=1,\ldots,m$. The complete state
space of the (partially) hidden process---now giving survival state
and covariate value---comprises these $m$ states plus the ``recent
dead'' (state $m+1$) and the ``long dead'' (state $m+2$) survival states.
To obtain the matrix product form of the likelihood, we extend the HMM
form described in Section~\ref{seclic1}, allowing for the augmentation
of the single alive state $s_{t}=1$ to the set of $m$ states. In
particular, we need to extend the definitions of the (system process)
matrix, $\bolds{\Gamma}_t$, observation matrix, $\mathbf{Q}_{t}$,
and an initial distribution for the covariate values, $\bolds
{\delta}$ (assuming that these are not always observed). First, we
define the $(m+2)\times(m+2)$ system process matrix
\[
\bolds{\Gamma}_{t}^{(m)} = \pmatrix{ {
\phi}_{t}(1)\Psi_t(1,1) & \cdots& {\phi}_{t}(1)
\Psi_t(1,m) & 1- {\phi }_{t}(1) & 0
\cr
\vdots & \ddots &
\vdots & \vdots & \vdots
\cr
{\phi}_{t}(m)\Psi_t(m,1) &
\cdots& {\phi}_{t}(m)\Psi_t(m,m) & 1- {\phi
}_{t}(m) & 0
\cr
0 & & \cdots & 0 & 1
\cr
0 & & \cdots & 0 & 1},
\]
where
\[
\Psi_t(i,j) = \cases{ f(y_{t+1} \vert y_{t}), &\quad
if $t, t+1 \in\mathcal{W}, y_{t} \in B_i,
y_{t+1} \in B_j$;
\vspace*{1pt}\cr
f\bigl(y_{t+1} \vert
b_i^*\bigr), &\quad if $t \in\mathcal{W}^c, t+1 \in\mathcal
{W}, y_{t+1} \in B_j$;
\vspace*{1pt}\cr
f(y_{t+1} \in
B_{j} \vert y_{t}), &\quad if $t \in\mathcal{W}, t+1 \in
\mathcal{W}^c, y_{t} \in B_i$;
\vspace*{1pt}\cr
f
\bigl(y_{t+1} \in B_{j} \vert b_i^*\bigr), &\quad if
$t,t+1 \in\mathcal{W}^c$;
\vspace*{1pt}\cr
0, &\quad otherwise,}
\]
and
\[
{\phi}_t(i) = \cases{ f(s_{t+1} = 1 \vert
s_{t}=1,y_{t}), &\quad if $t \in\mathcal{W}, y_t
\in B_i$;
\vspace*{1pt}\cr
f\bigl(s_{t+1} = 1 \vert s_{t}=1,b_i^*
\bigr), &\quad if $t \in\mathcal{W}^c$;
\cr
0, &\quad otherwise.}
\]
Here the product ${\phi}_{t}(i)\Psi_t(i,j)$ corresponds to the
probability of the individual surviving from time $t$ to time $t+1$,
with the covariate value changing from a given value in the interval
$B_i$ at time $t$ (either the observed covariate value or the
representative value) to some value in the interval $B_j$ at time $t+1$
(either the observed covariate value or any point within the interval).
We note that this formulation is similar to the Arnason--Schwarz model,
where the transition probabilities are defined between discrete states.
However, within our model specification the transition probabilities
are of a more complex form, as they are determined via the underlying
model specified on the covariate process (rather than estimated
freely), and also as they depend on whether the (continuous) covariate
value is observed or not. For example, the probability $f(y_{t+1} \in
B_{j} \vert b_i^*)$ is determined by the model used for the covariate
process. If the model given by (\ref{como}) is considered, then
\[
f\bigl(y_{t+1} \in B_{j} \vert b_i^*\bigr)=
\Phi \biggl( \frac{b_j- (b_i^*+a_{t})
}{\sigma} \biggr)-\Phi \biggl( \frac{b_{j-1}- (b_i^*+a_{t})}{\sigma} \biggr),
\]
where $\Phi$ denotes the cumulative distribution function of the
standard normal distribution.

We now consider the matrix comprising the state-dependent observation
probabilities, which is a diagonal matrix of dimension $(m+2)\times
(m+2)$, such that
\[
\mathbf{Q}^{(m)}(x_{t}) = \cases{\operatorname{diag}(1-p_{t},\ldots,1-p_{t},1-\lambda_{t},1), &\quad if $x_{t} =
0$;
\cr
\operatorname{diag}(p_{t},\ldots,p_{t},0,0), &\quad if
$x_{t} = 1$;
\cr
\operatorname{diag}(0,\ldots,0,\lambda_{t},0), &\quad
if $x_{t} = 2$.}
\]
Finally, one may need to model the initial distribution for the
covariate value (since the initial value may not be observed). In
general, the distribution will depend on the model assumed for the
covariate process. For a (conditional) probability density function of
initial covariate values given by $f_0$ (given the individual was
captured during the study), we define the row vector $\bolds{\delta
}^{(m)}$ of length $m+2$ with the $i$th element,
\[
\delta_i^{(m)} = \cases{\displaystyle \int_{b_{i-1}}^{b_i}
f_0(z) \,dz, &\quad if $g \in\mathcal{W}^c, i \in\{ 1,\ldots,m
\}$,
\vspace*{2pt}\cr
f_0 (y_g), &\quad if $g \in\mathcal{W},
y_g \in B_i$,
\vspace*{1pt}\cr
0, &\quad otherwise.}
\]
If all initial covariate values are observed and the initial covariate
distribution itself is not of interest, then one can set $\delta
_i^{(m)}=1$ for $g \in\mathcal{W}, y_g \in B_i$, which corresponds to
conditioning the likelihood on the initial covariate value (with the
advantage that less parameters have to be estimated).
Putting all these components together, the matrix formulation of (\ref
{lik4}) is
%
\begin{eqnarray}
\label{lik5}\quad
\mathcal{L} & = & \bolds{\delta}^{(m)}
\Biggl( \prod_{t=g+1}^{T} \bolds{
\Gamma}_{t-1}^{(m)} \mathbf{Q}^{(m)}(x_{t})
\Biggr) \mathbf{1}_{m+2}
\nonumber\\[-8pt]\\[-8pt]
& = & \bolds{\delta}^{(m)} \bolds{\Gamma}_{g}^{(m)}
\mathbf {Q}^{(m)}(x_{g+1}) \bolds{\Gamma}_{g+1}^{(m)}
\mathbf {Q}^{(m)}(x_{g+2}) \cdot\cdots\cdot\bolds{
\Gamma}_{T-1}^{(m)} \mathbf{Q}^{(m)}(x_{T})
\mathbf{1}_{m+2},\nonumber
\end{eqnarray}
that is, the likelihood has exactly the same structure as in the case
of absence of covariates [cf. expression (\ref{lik2})]. It should
perhaps be emphasized here that although (\ref{lik5}) has precisely the
same structure as an HMM likelihood (and hence can easily be maximized
numerically), it is not the likelihood of an HMM, since (for any given
$t$) the rows of the matrix $\bolds{\Gamma}_{t}^{(m)}$ in general
do not sum to one. This is because some of the covariate values are
known, and also because we restrict the range of covariate values to
some essential range.

\subsubsection{Inference}

For multiple individuals, the likelihood is simply the product of
likelihoods of type (\ref{lik5}), corresponding to each encounter
history. It is then a routine matter to numerically maximize this joint
likelihood with respect to the model parameters, subject to well-known
technical issues arising\vadjust{\goodbreak} in all optimization problems; see Chapter~3 in
\citet{zucm09} for a detailed account of the particular issues that
arise in the case of HMMs. Approximate confidence intervals for the
parameters can be obtained based on the estimated Hessian or,
alternatively, using a parametric bootstrap. Model selection, including
for the underlying covariate process model, can easily be carried out
using model selection criteria such as the Akaike information criterion (AIC).

The accuracy of the likelihood approximation increases with increasing
$m$. The influence on the estimates can be checked by considering
different values of $m$: if for some relatively large $m$ a further
increase does not change the likelihood value and/or the estimates,
then this is a very strong indication that $m$ is sufficiently large to
ensure a very close approximation. From our experience we suggest using
$20$--$80$ intervals in the discretization [cf. the simulation study in
\citet{lanmz12} and further remarks on this issue in Section~\ref{appl} below].

We note that the computational expense is not only a function of $m$
and $T$ and of the proportion of missing covariates, but also of the
pattern that the missing values occur in. Consecutive missing covariate
values lead to the highest computational burden (since they imply that
all entries of the corresponding system process matrix associated with
the underlying covariate process need to be calculated, a total of
$m^2$ entries). If an unobserved covariate value is followed by an
observed covariate value, then the corresponding system process matrix
consists of only one column with nonzero entries (and likewise, if an
observed covariate value is followed by an unobserved covariate value,
then there is only one row with nonzero entries). Consecutive observed
covariate values are clearly least computationally intensive (the
system process matrix then consists of only one nonzero element).

\section{Simulation study}
\label{simul}

In this section we present the results of a simulation study for
evaluating the performance of the HMM-based method. As a benchmark
method we consider the trinomal method suggested by \citet{catmt08},
which currently appears to be the most popular \textit{classical}
inference method for MRR models with continuous-valued covariates [\citet
{bonmk10}]. We considered four different simulation scenarios, using
different values for the recapture and the recovery probabilities,
respectively. Table~\ref{scen} gives the combinations of these
parameters that were considered. The different scenarios represent,
\textit{inter alia}, different amounts of information on the survival
states (the lower $\lambda$ and the lower $p$, the less information)
and on the covariate values (the lower $p$, the less information),
respectively. For each of the scenarios we conducted 500 simulation
experiments, in each experiment considering simulated capture histories
for $N=500$ individuals, each of them observed on at most $T=10$
occasions. For each individual the time of the initial capture occasion
was chosen uniformly from $\{1,\ldots,9\}$.

\begin{table}
\tablewidth=150pt
\caption{Configurations of true recovery and recapture probabilities
used in four different simulation scenarios}
\label{scen}
\begin{tabular*}{\tablewidth}{@{\extracolsep{\fill}}lcc@{}}
\hline
\textbf{Scenario} & $\bolds{p}$ & $\bolds{\lambda}$ \\
\hline
1 & 0.95 & 0.95 \\
2 & 0.90 & 0.30 \\
3 & 0.30 & 0.90 \\
4 & 0.30 & 0.30 \\
\hline
\end{tabular*}
\end{table}

In each scenario we used the same underlying process to generate the
covariate values. More precisely, for each individual we generated the
values of the covariate process using an autoregressive-type process of
order 1 with a deterministic (sine-shaped) trend:
\[
y_{t} - 25 = \eta(y_{t-1}-25) + \alpha_t +
\sigma\varepsilon_t,
\]
where $\alpha_t = \gamma\sin( 2\pi t/{T})$ and $\varepsilon_t \stackrel
{\mathrm{i.i.d.}}{\sim} \mathcal{N} (0,1)$. In all scenarios we used the following
values for the parameters determining the covariate process: $\eta
=0.6$, $\sigma=1.2$ and $\gamma=2$. For the initial (conditional)
covariate distribution, associated with the first capture event, we
used a normal with mean 15 and standard deviation 2. We assume a
logistic link function for the survival probabilities regressed on the
covariate values, with intercept $\beta_0=-3$ and slope $\beta_1=0.2$.
For this model the survival probability is $0.5$ for $w_{t-1}=15$ and
greater than $0.9$ for $w_{t-1}>26$. The parameter values were chosen
roughly similar to those estimated in the application to Soay sheep MRR
data given in \citet{bonmk10}. In particular, a typical covariate time
series starts at around 15 at the initial capture occasion, over the
years approaches 25 and then fluctuates around that value. The
deterministic trend $\alpha_t$ was included to enable us to conduct a
simple check for robustness of our method to model misspecification
(see below).

We here focus on the estimation of the parameters $\beta_0$ and $\beta
_1$, and in each case give the following summary statistics: sample
mean relative bias [$(\hat{\beta}_i-\beta_i)/\beta_i$], 2.5 and 97.5\%
quantiles of the relative bias, sample mean width of the estimated 95\%
confidence intervals and coverage probability of the confidence
intervals. Confidence intervals were obtained based on the estimated
Hessian matrix. For the HMM-based method we considered three different
covariate process models in the simulation experiments: (1) the
correctly specified model (i.e., the one that was used for simulating
the data; model HMM-C), (2) a slightly misspecified model which assumes
a homogeneous AR(1) for the covariate process (i.e., one that neglects
the deterministic sine-shaped component of the trend; model \mbox{HMM-M1}),
and (3) a substantially misspecified model which assumes that at all
ages (and across all individuals) the covariate is independently and
identically normally \mbox{distributed}, with mean and standard error being
estimated in the simulation experiments (model HMM-M2; this model
neglects both trend components and correlation over time). The latter
two explore the robustness of our method to misspecification of the
covariate process model. In the implementation of our approach we used
$m=40$ intervals in the discretization of the covariate space, and the
function \texttt{nlm} in R to maximize the approximate likelihood
numerically. In the implementation of the trinomial approach we used
the function \texttt{optim} in~R instead, since \texttt{nlm} had
problems in estimating the Hessian when $p$ or $\lambda$ are estimated
at the boundaries of their support (which happens occasionally when
using the trinomial method). Sample \texttt{R} code for simulating data
and fitting the corresponding model using the HMM-based approach is
given in the supplementary material [\citet{lan13}]. Results are
provided in Table~\ref{simresults}.

%
%
\begin{table}
\tabcolsep=0pt
\caption{Sample means and 2.5 and 97.5\% quantiles of the relative
biases (RB), sample mean widths (CW) of the estimated 95\% confidence
intervals and coverage probabilities (CC) of the confidence intervals,
for the logistic regression parameters $\beta_0$ and $\beta_1$, in four
simulation scenarios}
\label{simresults}
\begin{tabular*}{\tablewidth}{@{\extracolsep{4in minus 4in}}lclcccccc@{}}
\hline
& & & \multicolumn{3}{c}{\textbf{Intercept ($\bolds{\beta_0=-3}$)}} &
\multicolumn
{3}{c@{}}{\textbf{Slope ($\bolds{\beta_1=0.2}$)}} \\[-4pt]
& & & \multicolumn{3}{c}{\hrulefill} & \multicolumn{3}{c@{}}{\hrulefill
} \\
\textbf{Scenario} & \textbf{Meth.} & & $\bolds{\operatorname
{RB}(q_{0.025}, q_{0.975})}$
& \textbf{CW} & \textbf{CC} & $\bolds{\operatorname{RB}(q_{0.025}, q_{0.975})}$
& \textbf{CW} & \textbf{CC} \\
\hline
1 & Tri & & \hphantom{$-$}0.00 ($-$0.23, 0.22) & 1.39 & 0.96 & \hphantom
{$-$}0.00 ($-$0.20, 0.20) & 0.08 & 0.94 \\
& HMM-C & & \hphantom{$-$}0.00 ($-$0.24, 0.23) & 1.33 & 0.94 & \hphantom
{$-$}0.00 ($-$0.19, 0.21) & 0.07 & 0.93 \\
& HMM-M1 & & \hphantom{$-$}0.00 ($-$0.24, 0.22) & 1.34 & 0.94 & \hphantom
{$-$}0.00 ($-$0.18, 0.21) & 0.08 & 0.94 \\
& HMM-M2 & & $-$0.01 ($-$0.24, 0.22) & 1.34 & 0.95 & \hphantom
{$-$}0.01 ($-$0.19, 0.21) & 0.08 & 0.94 \\[4pt]
2 & Tri & & \hphantom{$-$}0.00 ($-$0.28, 0.26) & 1.69 & 0.95 & \hphantom
{$-$}0.00 ($-$0.27, 0.30) & 0.12 & 0.95 \\
& HMM-C & & \hphantom{$-$}0.00 ($-$0.24, 0.22) & 1.37 & 0.95 & \hphantom
{$-$}0.00 ($-$0.18, 0.20) & 0.08 & 0.95 \\
& HMM-M1 & & \hphantom{$-$}0.00 ($-$0.24, 0.21) & 1.38 & 0.96 & \hphantom
{$-$}0.00 ($-$0.18, 0.20) & 0.08 & 0.95 \\
& HMM-M2 & & $-$0.03 ($-$0.28, 0.20) & 1.41 & 0.95 & \hphantom
{$-$}0.02 ($-$0.17, 0.23) & 0.08 & 0.94 \\[4pt]
3 & Tri & & \hphantom{$-$}0.03 ($-$0.52, 0.34) & 3.08 & 0.97 & \hphantom
{$-$}0.02 ($-$0.26, 0.35) & 0.14 & 0.97 \\
& HMM-C & & \hphantom{$-$}0.00 ($-$0.26, 0.24) & 1.46 & 0.94 & \hphantom
{$-$}0.00 ($-$0.21, 0.22) & 0.08 & 0.95 \\
& HMM-M1 & & \hphantom{$-$}0.00 ($-$0.27, 0.26) & 1.50 & 0.93 & \hphantom
{$-$}0.01 ($-$0.21, 0.24) & 0.09 & 0.94 \\
& HMM-M2 & & \hphantom{$-$}0.01 ($-$0.28, 0.28) & 1.57 & 0.93 &
$-$0.01 ($-$0.23, 0.23) & 0.09 & 0.94 \\[4pt]
4 & Tri & & \hphantom{$-$}0.00 ($-$0.58, 0.60) & 3.73 & 0.98 & \hphantom
{$-$}0.01 ($-$0.45, 0.57) & 0.20 & 0.95 \\
& HMM-C & & \hphantom{$-$}0.00 ($-$0.30, 0.32) & 1.92 & 0.95 & \hphantom
{$-$}0.00 ($-$0.26, 0.25) & 0.11 & 0.95 \\
& HMM-M1 & & $-$0.01 ($-$0.34, 0.33) & 2.01 & 0.95 & \hphantom
{$-$}0.02 ($-$0.26, 0.31) & 0.11 & 0.95 \\
& HMM-M2 & & $-$0.09 ($-$0.45, 0.30) & 2.20 & 0.92 & \hphantom
{$-$}0.09 ($-$0.23, 0.40) & 0.13 & 0.92 \\[4pt]
\hline
\end{tabular*}
\end{table}

In all four simulation scenarios, the interval estimates obtained using
the HMM-based method were narrower than those obtained using the
trinomial method, with the differences being substantial in scenarios 3
and 4 (those with low capture probabilities). Using the HMM-based
method, with both the correct specification (HMM-C) and with a slight
misspecification (HMM-M1) of the model for the covariate process, no
significant bias was found in the estimates of the logistic regression
parameters (for each scenario). The experiment involving a substantial
misspecification of the covariate process model (HMM-M2) led to a 9\%
negative bias in scenario 4 (with both low capture and recovery
probabilities), whereas in all other scenarios there still was only a
small bias. In all considered settings, coverage probabilities of the
interval estimates were close to 95\%. We note that it is immediate to
consider a model selection approach for the underlying covariate
process, for example, using the AIC statistic. For the present
simulation experiment, the correct underlying covariate model (model
HMM-C) was deemed optimal by the AIC statistic in all 500 simulation
runs (when compared to the models HMM-M1 and HMM-M2, resp.). 

We conclude this section with some remarks on the computing times
involved. On an octa-core i7 CPU, at 2.7 GHz and with 4 GB RAM, the
simulation runs took, on average per run, 15, 18, 14 and 15 seconds for
scenarios 1, 2, 3 and 4, respectively, when applying the trinomial
method, and 3, 15, 8 and 20 minutes for the same scenarios when
applying the HMM method (with the correct model specification and
$m=40$). The computational effort is thus extremely low for
the~trinomial method and modest for the HMM approach (for reasonable
$m$). In the case of the HMM approach, the computational effort is
highly dependent on the desired accuracy of the likelihood
approximation: for example, in scenario 3, the average computing time
per simulation run is 2 minutes when using $m=10$ intervals in the
discretization and 54 minutes when using $m=150$.

\section{Application to Soay sheep data}
\label{appl}

We consider capture histories for Soay sheep that were born and tagged
on the Island of Hirta, off the west coast of Scotland, between 1985
and 2009, with the annual surveys being carried out in the summer.
These sheep have been the subject in numerous studies on population
dynamics, due to their isolated nature with no natural
predators---Hirta was left by the last residents in 1932, after which
the sheep established a wild population---and the ease with which
individuals can be marked and recaptured. Annual studies involving,
\textit{inter alia}, captures, searches for dead animals and weighings are
conducted. We consider only female sheep, with at least one recorded
body mass, leading to a total of 1344 individual capture histories. The
mean number of observations per sheep is 4.64, with a total of 900
sheep recovered dead during the observation period. We assume that the
survival probabilities are a function of body mass, noting that the
primary cause of mortality is starvation, with the risk of dying from
starvation being highest for young individuals. It is not the objective
of the given analysis to perform a full investigation of the factors
that affect\vadjust{\goodbreak} the survival of the individuals. For details on the
population dynamics of the Soay sheep we refer to \citet{clup04}.

Not all observations are associated with the animal being physically
captured, and thus for 38\% of the observations the corresponding body
mass was not recorded. Following \citet{bonmk10}, we consider four
different age groups: lambs ($\mbox{age} <1$), yearlings ($\mbox{age} \in[1,2)$),
adults ($\mbox{age} \in[2,7)$) and seniors ($\mbox{age} \ge7$). We assume a
logistic relationship between the covariate body mass and the survival
probability, so that
\[
\operatorname{logit} ( \phi_{t} ) = \beta_{a_{t},0} +
\beta_{a_{t},1} y_{t}.
\]
For a given sheep, $a_{t}$ indicates the age group the sheep is in at
time $t$ (lamb, yearling, adult or senior). We consider five different
possible models in total, summarised as follows:
\begin{longlist}[Model 5:]
\item[Model 1:] $y_{t}= y_{t-1} + \eta_{a_{t}}  ( \mu_{a_{t}} -
y_{t-1}  ) + \sigma_{a_{t}} \varepsilon_{t}$ (i.e., distinct
covariate process parameters across age groups), time-dependent
recapture probabilities, time-dependent recovery probabilities (68 parameters);
\item[Model 2:] $y_{t}= y_{t-1} + \eta ( \mu- y_{t-1}  ) +
\sigma\varepsilon_{t}$ (i.e., covariate process parameters fixed across
age groups), time-dependent recapture probabilities, time-dependent
recovery probabilities (59 parameters);
\item[Model 3:] Same covariate model as for model 1, constant recapture
probability, constant recovery probability (22 parameters);
\item[Model 4:] Same covariate model as for model 1, constant recapture
probability, time-dependent recovery probability (45 parameters);
\item[Model 5:] $y_{t}= y_{t-1} + \mu_{a_{t}} + \sigma_{a_{t}} \varepsilon
_{t}$, time-dependent recapture probabilities, time-dependent recovery
probabilities (64 parameters).
\end{longlist}
For each model, $\varepsilon_{t}$ denote independently and identically
distributed standard normal random variables. Model 5 has a covariate
process model similar to those used by \citet{bons06}, \citet{kinbc08}
and \citet{bonmk10} [although, e.g., \citet{bonmk10} assume $\mu$
not only depends on the age group of the sheep but also on the year and
\citet{kinbc08} consider a further additive year effect]. Notably, this
covariate process model is diffusive and thus, in general, not
biologically realistic (see later discussion). 

\begin{table}
\tablewidth=150pt
\caption{Log-likelihood, number of parameters ($q$) and $\Delta$AIC
values for different models, fitted to the Soay sheep data}
\label{est2}
\begin{tabular*}{\tablewidth}{@{\extracolsep{\fill}}lccc@{}}
\hline
& $\bolds{\log\mathcal{L}}$ & $\bolds{q}$ & $\bolds{\Delta}$\textbf{AIC} \\
\hline
Model 1 & $-$10\mbox{,}222 & 70 & \hphantom{00}0 \\
Model 2 & $-$10\mbox{,}351 & 61 & 240 \\
Model 3 & $-$10\mbox{,}309 & 24 & \hphantom{0}83 \\
Model 4 & $-$10\mbox{,}261 & 47 & \hphantom{0}32 \\
Model 5 & $-$10\mbox{,}405 & 66 & 357 \\
\hline
\end{tabular*}
\end{table}

\begin{figure}[b]

\includegraphics{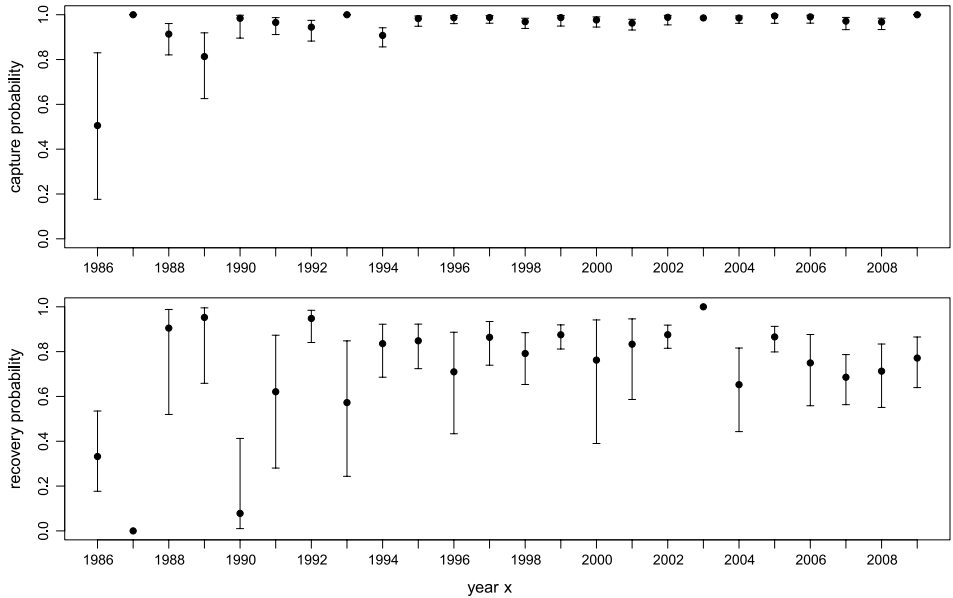}

\caption{Estimates of the yearly recapture and recovery probabilities
obtained for model 1. Points represent the ML estimates, and error bars
indicate the 95\% confidence intervals (only for those estimates that
do not lie at the boundary of the parameter space).} \label{caprecap}
\end{figure}

Each of the models was fitted using the HMM-based approach using $m =
50$ intervals in the discretization. The assumed essential range of
covariate values is given by $b_0 = 0.8 b_{\mathrm{min}}$ and $b_m = 1.2
b_{\mathrm{max}}$, where $b_{\mathrm{min}}$ and $b_{\mathrm{max}}$ denote the minimum and the
maximum of the observed covariate values, respectively. For the given
data, $b_{\mathrm{min}} = 2.9$ and $b_{\mathrm{max}} = 33.9$. For the initial covariate
value we assumed a normal distribution and estimated the corresponding
mean and variance parameter alongside the other parameters. For the
different models considered, the computing time ranged from 14 hours
(for model 3) to 45 hours (for model 1); the computing times are much
higher than those observed in the simulation experiments described in
Section~\ref{simul}, which is primarily due to the high-dimensional
parameter spaces associated with the models fitted to the real data.
The log-likelihood and $\Delta$AIC values obtained for the five
different models described above are provided in Table~\ref{est2}.
Clearly, model 1 is identified as optimal via the AIC statistic by
quite a substantial margin.

Figure~\ref{caprecap} displays the estimated year-dependent recapture
and recovery probabilities for model 1. The results generally match
those of \citet{bonmk10} well for the years common between the analyses
(i.e., 1986--2000), except in the initial two years. This mismatch
appears to be related to the use of slightly different data: for
example, in our data set there are no recoveries in 1987, but \citet
{bonmk10} estimate a positive recovery probability in that year. The
variability over time in the recovery probabilities is considerably
greater than for the recapture probabilities, which is also identified
via the model selection procedure above (see $\Delta$AIC values in
Table~\ref{est2}). 

\begin{figure}

\includegraphics{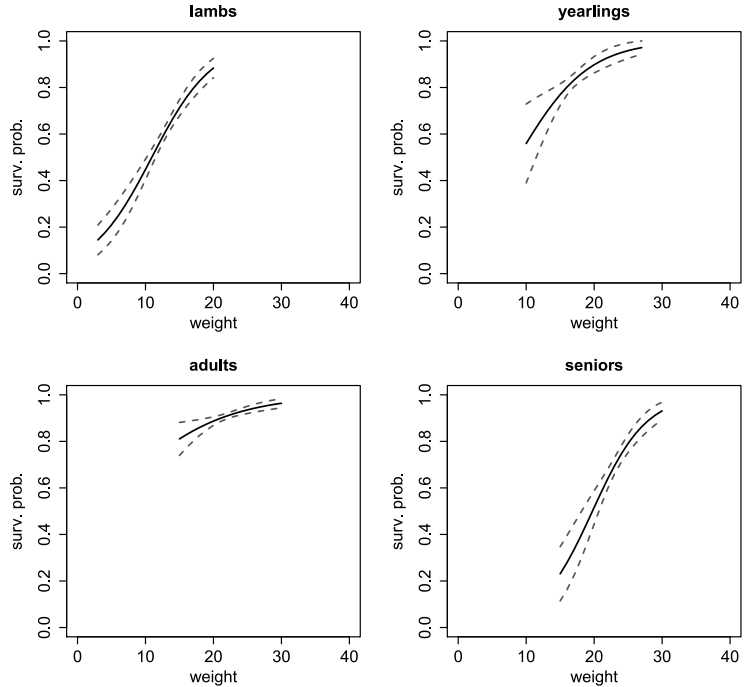}

\caption{Estimated survival probability as a function of the covariate
body mass (in kg), for the four different age groups (for model 1).
Solid lines give the maximum likelihood estimates, and dashed lines
indicate the 95\% pointwise confidence intervals.} \label{survivals}
\end{figure}

Figure~\ref{survivals} displays the estimated survival probabilities
for model 1 for the different age groups, in each case as a function of
body mass. Pointwise confidence intervals were obtained based on the
Hessian (via the delta method). Again, the results are similar to those
of \citet{bonmk10}. The survival probability increases with increasing
body mass, with this effect found strongest for lambs and seniors, and
weakest for adults. The interval estimates are slightly narrower than
those obtained by \citet{bonmk10}, which is not surprising given that we
consider a larger data set.\looseness=-1

\begin{figure}

\includegraphics{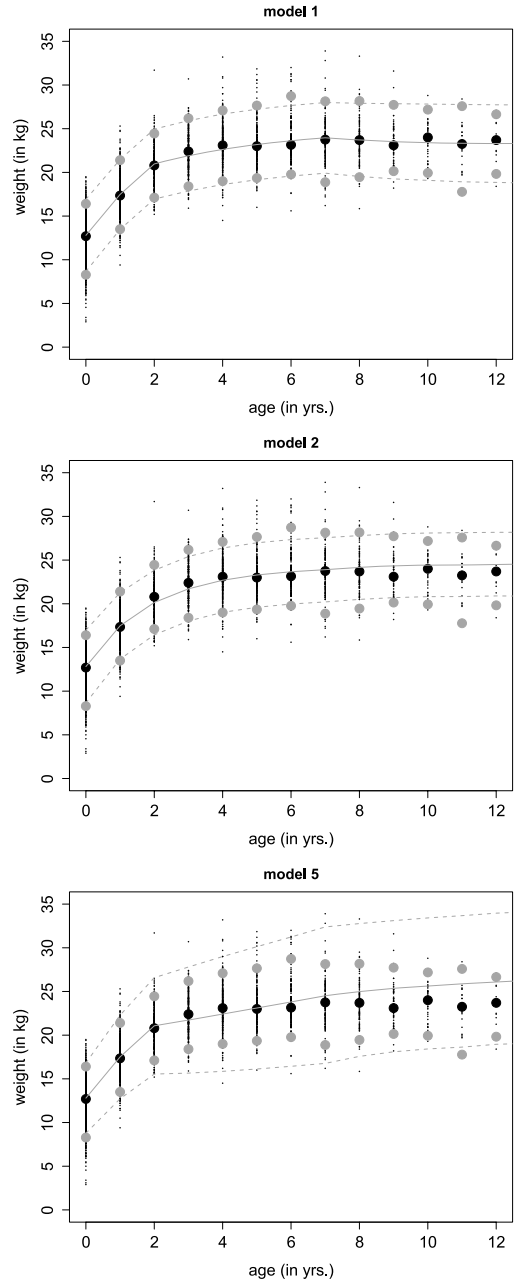}

\caption{Observed body masses of sheep at ages 0--12 (tiny black dots),
empirical 5\% and 95\% quantiles (big grey dots) and empirical medians
(big black dots) of body masses at those ages, and model-derived 5\%
and 95\% quantiles (dashed grey lines) and medians (solid black lines)
of body mass distributions of alive individuals at those ages (obtained
through simulation), for fitted models 1, 2 and 5.}
\label{weightprocess}
\end{figure}

In Figure~\ref{weightprocess}, the observed body masses of sheep at
ages 0--12 are compared to the model-derived distributions of body
masses (of alive sheep) for these ages. We omitted models 3 and 4 since
the covariate process model in these models is identical to that of
model 1. Models 1 and 2 appear to capture the development of the body
mass over the years. However, the diffusive nature of the covariate
process in model 5 leads to increasingly wider interval estimates for
body mass as age increases, with the intervals not capturing well the
observed quantiles. Thus, as already identified via the AIC statistic,
it appears that the nondiffusive, auto-regressive type covariate
process models are more appropriate in this application.

\begin{figure}

\includegraphics{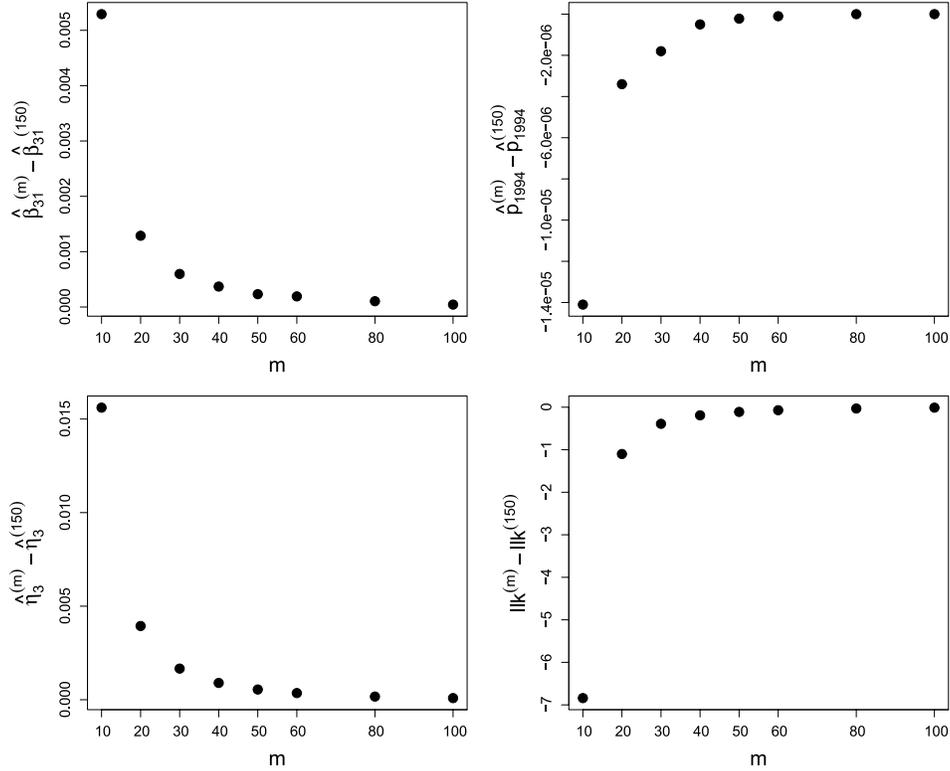}

\caption{Approximation error arising from the
discretization: differences between the estimates
$\hat{\beta}_{3,1}^{(m)}$, $\hat{p}_{1994}^{(m)}$ and
$\hat{\eta}_3^{(m)}$ (for given $m$, with
$m=10,20,30,40,50,60,80,100$), respectively, and the corresponding
estimates obtained for $m=150$ [$\hat{\beta}_{3,1}^{(150)}=0.122$;
$\hat{p}_{1994}^{(150)}=0.908$; $\hat{\eta}_3^{(150)}=0.222$], and
differences between the log-likelihood value for given $m$ and the
log-likelihood value obtained using $m=150$
($\mathrm{llk}^{(150)}=-10\mbox{,}221.74$).} \label{errorm}
\end{figure}

Finally, to investigate the effect of the choice of $m$, the number of
intervals used in the numerical integration of the likelihood, we
repeatedly ran the estimation of model 1, for
$m=10,20,30,40,50,60,80,100,150$. Figure~\ref{errorm} illustrates,
exemplarily, the convergence of the estimates $\hat{\beta
}_{3,1}^{(m)}$, $\hat{p}_{1994}^{(m)}$ and $\hat{\eta}_3^{(m)}$ (with
the superscript indicating their dependence on $m$) as $m$ increases
and also the convergence of the corresponding log-likelihood. In this
application, $m=50$ seems to provide a reasonable compromise between
minimizing the computational effort and maximizing the accuracy of the
numerical integration (although we note that even for $m=20$ the
estimates are close to those obtained for $m=150$). Not surprisingly,
the effect of the choice of $m$ is found to be strongest on the
estimates of parameters that are related to the covariate process (in
our example, $\hat{\eta}_3^{(m)}$), and weakest on the estimates of
parameters related to the observation process (here, $\hat{p}_{1994}^{(m)}$).

\section{Discussion}
\label{discuss}

In recent years, several different methods have been proposed that
address MRR studies that involve individual-specific and time-varying
continuous covariates [see \citet{catmt08} for a summary of these
approaches]. The most popular approaches for fitting models to this
type of data are the conditional trinomial method [\citet{catmt08}] and
the Bayesian imputation method [\citet{bons06,kinbc08,kinmgb09,schb11}].
The former method is easy to implement, computationally fast and avoids
assumptions concerning the underlying model for the covariate process.
However, it disregards a potentially significant amount of information
in the data, which can lead to poor precision of the parameter
estimates. Use of the trinomial approach is not recommended if capture
probabilities are low [\citet{bonmk10}] or, clearly, if the underlying
covariate process is of interest in itself. While the Bayesian approach
is much more computer intensive than the trinomial method, it makes use
of all available information in the data and thus usually leads to an
improved precision of the estimators (provided a correct specification
of the covariate process model). However, prior distributions need to
be specified on all model parameters, and model selection is generally
more difficult and potentially sensitive to the prior specification.

The proposed HMM-based method for estimating such MRR models is based
on a discretization of the space of covariate values, which reduces the
multiple integral appearing in the likelihood to a multiple sum. The
resulting multiple sum can efficiently be calculated by rewriting it as
a matrix product that corresponds to a recursive scheme for evaluating
the (approximate) likelihood. While the fitting is based on maximizing
only an approximation to the likelihood, it is very easy to make this
approximation extremely accurate (by considering increasingly finer
discretizations of the covariate space), while maintaining
computational tractability in typical MRR settings. The HMM method is
fairly easy to implement and to apply [R code is provided in \citet
{lan13}] and, once it is implemented, changes of the model structure
usually only require very minor and straightforward changes to the
code, making this method very user-friendly.

The simulation study demonstrated that if the covariate process is
modelled adequately, and even if the model is misspecified to some
degree, then the HMM-based approach leads to more precise estimates
than does the trinomial method. The difference in the precision is
small if (and only if) there are only few missing covariate values, and
in such a case the trinomial method can be more attractive due to the
extremely low computational effort it involves, and as it is
implemented in the widely used software package MARK [\citet{bon12}].
If, however, the covariate process is also of interest, then the HMM
method has the additional advantage of allowing for formal (and simple)
comparison between competing covariate process models (using standard
information criteria). Model checking of the covariate process model
can be performed by comparing the observed covariate values with those
obtained from the fitted process model, for example, using graphical
means to assess a lack of model fit. 

We applied the novel HMM-based approach to MRR data collected on female
Soay sheep born between 1985 and 2009, investigating the effect of body
mass on survival and comparing a variety of models for the change of
the covariate body mass over time. Previous covariate process models
that have been suggested for these type of data (including the Soay
sheep) are typically of the form of diffusive random walks. For this
application, an alternative nondiffusive AR(1)-type model appears to
provide a significantly better fit, particularly at increasing age of
the sheep (which is due to the model-derived variance of body mass
diverging as age increases in the case of the diffusive random walk).
The AR(1)-type model is similar to, but more flexible than, the von
Bertalanffy growth curve model [\citet{jam91}], distinguishing between
different age classes within which growth, or change of body mass (in
the Soay sheep application), is homogeneous. We believe that this type
of model has the potential to be very useful for analyzing
growth-related dynamics. The results obtained showed an increasing
survival probability with increasing body mass for each age group. The
strongest effect was observed for lambs and seniors, and the weakest
for adults, corresponding well to findings of previous studies [\citet
{bonmk10}]. This is biologically sensible, with the youngest and the
oldest sheep the ``weakest'' individuals and less able to compete for
available resources. Recapture probabilities were estimated to vary
only slightly over time, while the estimated recovery probabilities
showed great variability over time. Further research involves the
consideration of multiple covariates (see below) and different
age-dependence structures to identify further biological structure.

The HMM-based approach can be extended in different ways. The extension
to allow the observation model parameters to be dependent on the
individual covariate is straightforward, with minimal additional
computational effort---the only change that is required relates to the
matrix comprising the state-dependent probabilities. A drawback of the
HMM-based approach is that the computational effort increases
dramatically if \textit{multiple} continuous, individual-specific and
time-varying covariates are considered, and in such cases a Bayesian
approach will often be preferable. However, we anticipate that using
more sophisticated numerical procedures in the likelihood
approximation, such as, for example, Gauss--Legendre, will at least
render the case of two such covariates feasible even for relatively
large MRR data sets. In general, it may also be worthwhile to consider
alternative numerical approaches for evaluating the likelihood, such
as, for example, simulated maximum likelihood [which is often used in
stochastic volatility modelling; see, e.g., \citet{duko}]. Another
extension that is straightforward in principle, but accompanied by
large scale increases in computational time, is that to models
involving random effects [see, e.g., \citet{kinbc08} for an account in
an MRR setting in a Bayesian framework, and \citet{schlIF} and \citet
{langEC} for implementations of similar models in a non-Bayesian HMM
framework in other ecological applications].

\section*{Acknowledgements}

We would like to thank all the volunteers and members of the Soay sheep
project who collected the data from the sheep at St Kilda, and the
National Trust for Scotland and the Scottish Natural Heritage who have
supported this project. In addition, we would like to thank Tim Coulson
for providing the data and for discussions with regard to the data,
Hannah Worthington for preprocessing the data, and Rachel McCrea and
Byron Morgan for valuable comments.

\begin{supplement}
\stitle{R code for model fitting}
\slink[doi]{10.1214/13-AOAS644SUPP} 
\sdatatype{.txt}
\sfilename{aoas644\_supp.txt}
\sdescription{Sample \texttt{R} code for simulating MRR data and
fitting the corresponding model using the HMM-based approach (with MRR
model as described in Section~\ref{simul}).}
\end{supplement}


\printaddresses

\end{document}